\documentclass[]{aa}  
\usepackage{graphicx}
\usepackage{txfonts}
\usepackage{color}

\def\compps{{\sc compps}}

\def\Integ{{\em INTEGRAL}} 
\def\BS{{\em Beppo-SAX}} 
\def\rxte{{\em RXTE}} 
\def\swift{{\em Swift}} 
 
\def\chandra{{\em Chandra}} 
\def\xmm{{\em XMM-Newton}} 
\def\sax{SAX~J1748.9--2021} 
\def\be{\begin{equation}} 
\def\ee{\end{equation}}

\begin{document} 

\title{Mixed H/He bursts in SAX J1748.9--2021 during the spectral change of its 2015 outburst} 

\titlerunning{Mixed H/He bursts in SAX J1748.9--2021 during its 2015 outburst}  
\authorrunning{Li et al.}  
 
\author{Z. Li\inst{1,2,3}
\and V. De Falco\inst{4,2}
\and M. Falanga\inst{2,5}
\and E. Bozzo\inst{6}
\and L. Kuiper\inst{7} 
\and J. Poutanen\inst{8,9,10}  
\and A. Cumming\inst{11}
\and D. K. Galloway\inst{12,13}
\and S. Zhang\inst{14}      
} 
 
%\offprints{Z. Li} 
 
\institute{Department of Physics, Xiangtan University, Xiangtan, 411105, P. R. China \\
\email{lizhaosheng@xtu.edu.cn} 
\and International Space Science Institute (ISSI), Hallerstrasse 6, 3012 Bern, Switzerland
\and  Albert Einstein Center for Fundamental Physics, Institute for Theoretical Physics / Laboratory for High-Energy Physics, University of Bern, Switzerland
\and Departement Physik, Universit\"at Basel, Klingelbergstrasse 82, 4056 Basel, Switzerland
\and International Space Science Institute Beijing, No.1 Nanertiao, Zhongguancun, Haidian District, 100190 Beijing, China
\and Department of Astronomy, University of Geneva, chemin d'ecogia 16, 1290, Versoix, Switzerland
\and SRON, Netherlands Institute for Space Research, Sorbonnelaan 2, 3584 CA, Utrecht, The Netherlands 
\and Tuorla Observatory, Department of Physics and Astronomy,  FI-20014  University of Turku, Finland
\and Space Research Institute of the Russian Academy of Sciences, Profsoyuznaya str. 84/32, 117997 Moscow, Russia 
\and Nordita, KTH Royal Institute of Technology and Stockholm University, Roslagstullsbacken 23, SE-10691 Stockholm, Sweden
\and Department of Physics and McGill Space Institute, McGill University, 3600 Rue University, Montreal QC, Canada H3A 2T8
\and School of Physics and Astronomy, Monash University, VIC 3800, Australia 
\and Monash Centre for Astrophysics, Monash University, VIC 3800, Australia 
\and Laboratory for Particle Astrophysics, Institute of High Energy Physics, Beijing 100049, P. R. China
} 
 
\date{Accepted for publication in A\&A on 12/10/2018 } 
 
\abstract{
SAX~J1748.9--2021 is a transiently accreting X-ray millisecond pulsar. It is also known as an X-ray burster source discovered by \BS.  We analysed the persistent emission and type-I X-ray burst properties during its 2015 outburst. The source varied from hard to  soft state within half day. We modeled the broad-band spectra of the persistent emission in the (1 -- 250) keV energy band for both spectral states using  the quasi-simultaneous {\em INTEGRAL} and {\em Swift} data.  The broad-band spectra are well fitted by an absorbed thermal Componization model, {\sc compps}, in a slab geometry. The best-fits  for the two states indicate significantly different plasma temperature of 18 and 5 keV and the Thomson optical depth of 3 and 4, respectively.
In total,  56 type-I X-ray bursts were observed during the 2015 outburst, of which 26 detected by {{\em INTEGRAL}} in the hard state,  25 by {{\em XMM-Newton}} in the soft state, and 5 by {{\em Swift}} in both states. As the object transited from  the hard to the soft state, the recurrence time for X-ray bursts decreased from  $\approx 2$ to $\approx1$ hr. The relation between the recurrence time, $ \Delta t_{\rm rec}$, and the local mass accretion rate per unit area onto the compact object, $\dot m$, is fitted by a power-law model, and yielded as best fit at $\Delta t_{\rm rec} \sim \langle  \dot{m} \rangle^{-1.02\pm0.03}$ using all X-ray bursts. In both cases, the observed recurrence times are consistent with the mixed hydrogen/helium bursts. We also discuss the effects of type-I X-ray bursts prior to the hard to soft transition.
}
 
%In the soft state, the source emitted X-ray photons as high as 200 keV, while in the hard soft, the X-ray photon energy truncated at $\sim$45 keV.  
%We analyzed the type-I X-ray bursts from \sax during its 2015 outburst.  

\maketitle

\section{Introduction}  
\label{sec:intro}

Neutron stars (NSs), in low mass X-ray binaries (LMXBs), accrete matter from their companion stars, forming  an accretion disk \citep{frank02}. The accreted matter on the NS surface can trigger thermonuclear flashes, called type-I X-ray bursts  \citep[see e.g.,][]{Lewin93}. The burst spectra are described by a blackbody with peak temperatures reaching $kT_{\rm bb}\approx 3$ keV, and with a gradual softening due to the cooling of the NS photosphere \citep[see][for a review]{Lewin93,strohmayer06,Galloway08}. The burst peak luminosity can reach the Eddington limit, $L_{\rm Edd}\approx 2\times10^{38}$ erg s$^{-1}$, and the total burst energy release is of the order $\sim 10^{39-42}$ erg, depending on the burst type \citep[e.g.,][]{Galloway08}. 

During type-I X-ray bursts, the injected soft X-ray photons may affect the NS hot inner accretion flow or the surrounding hot corona. Indeed, a deficit has been observed at 30 -- 50 keV photons in IGR J17473--2721, Aql X-1, 4U 1636--536, and GS 1826--238, implying that the soft X-ray photons, emitted during type-I X-ray bursts, cool down the electrons producing hard X-ray emission \citep{Chen12, Chen13, Ji13, Ji14a, ji15}. The hard persistent X-ray component is thought to be derived from  Comptonization in the hot corona \citep{Chen12,Chen13}. %\red{JP: I removed the following sentence, I find it incomprehensible: However, the corona is reheated on short timescales within the burst duration time requiring a large amount of energy, that might come from an enhancement of the accretion rate \citep{walker92},  from the disk evaporation \citep{Ji14b}, or from  magnetic reconnection \citep{Liu02}.}  
A similar study has also been performed during the hard state of the source 4U 1728--34, however, during the occurrence of type-I X-ray bursts the persistent hard X-ray deficit was not detected by {\em Rossi X-ray Timing Explorer}, due to the limitation of the response in hard X-ray band \citep{Ji14b}, but confirmed by {\em INTEGRAL} \citep{Kajava17}. The persistent emission level above 40 keV was about a factor of three lower compared with the persistent emission without burst activities \citep{Kajava17}. Also in this case the interpretation indicates that the soft radiation injection into the accretion disk or corona may induce spectral variability during type-I X-ray bursts. 

It is believed that with varying mass accretion rate  the spectral changes of NSs in LMXBs follow an atoll or $Z-$track on an X-ray color-color diagram (CCD) \citep{hasinger89,Klis06}. The accretion rate  increases
from the hard  (island) spectral state, where the soft emission is much reduced and the spectrum is dominated by a power-law-like spectrum up to the energies of
200 keV, to the soft (banana) spectral state, where most of the energy is emitted below $\sim 20$ keV \citep[e.g.,][]{Barret01,Gierlinski02}. The spectral change has been studied during type-I X-ray bursts \citep{Falanga06}, while  we  study \sax, as an example to investigate  
if the large number of type-I X-ray bursts during its 2015 February -- June  outburst can trigger a spectral state change, i.e., from a persistent hard to a soft spectral state.  In this paper, we report on the \Integ\ and \swift\ observations of \sax. The properties of the largest set of X-ray bursts as well as the correlation of all observed type-I X-ray bursts with the source in the soft and hard state are investigated. 

%We study \sax, as an example to investigate  if the large number of type-I X-ray bursts during its 2015 February -- June  outburst can trigger a spectral state change.
%The CCDs are powerful instruments of parameterizing the spectral changes using physically motivated spectral models to understand the underlying physical changes in the source emission. 
\subsection{The source SAX~J1748.9--2021}  
\label{sec:source}

The X-ray source \sax\ was discovered by \BS\ during an outburst  in 1998  \citep{Intzand98}. The source has been identified to be located within the globular cluster NGC~6440 at a distance of $\approx 8.2$ kpc \citep[][and references therein]{Intzand99,Valenti07}. The detection of a type-I X-ray burst associated to \sax\ confirmed the source as a NS hosted in a  LMXB system \citep{Intzand01}. \sax\ has also been observed during its quiescent state with \chandra\, that identified its optical counterpart candidate \citep[][and references therein]{Intzand01}.  \citet{Altamirano08} suggested that the companion star might be a main-sequence or a slightly evolved star with a mass ranging between (0.85 -- 1.1) $M_\odot$. Recently, the companion star has been confirmed by the Hubble Space Telescope, and it turns out to be  a main-sequence star with a mass of (0.70 -- 0.83) $M_{\odot}$, a radius of $(0.88\pm0.02)~R_{\odot}$, and a surface temperature of $(5250\pm80)~{\rm K}$ \citep{cadelano17}. This source has been observed in outburst in 1998, 2001, 2005, 2009 -- 2010, and 2015 and classified also as an atoll source \citep[][]{Intzand98, Intzand01, Markwardt05, Suzuki09,Patruno09,Bozzo15,Sanna16,Pintore16}. During both 2001 and 2005 outbursts the source showed intermittent pulsations at 442 Hz on timescales of hundreds of seconds \citep{Graviil07,Altamirano08}, from which it was possible to infer the orbital period of $\approx8.76$~hr and the magnetic field of $B \stackrel{>}{{}_\sim} 1.3\times 10^8$~G \citep{Altamirano08}. Therefore, this source has been classified as an accreting millisecond X-ray pulsar \citep[AMXP; see][for reviews]{Poutanen06,patruno12}.
% \citet{Guver13}, using the time resolved spectroscopy of the thermonuclear X-ray bursts from \sax, attempted to infer mass and radius estimated to be $R\approx8.2$~km and $M\approx1.78$~M$_\odot$ or $R\approx11$ km and $M\approx1.33$~M$_\odot$. 

% a strong H$\alpha$ emission from

\section{Observations and data}
\subsection{INTEGRAL}
\label{sec:integral} 

The data were obtained with \Integ\ \citep{w03} from 2015 February 16 to April 20, i.e., from satellite revolution  1508 to 1532 for a total effective exposure time of $\sim856$ ks. These revolutions include a public Target of Opportunity (ToO) observation for $\sim100$~ks during the revolution 1511.

We analyzed data from the coded mask imager IBIS/ISGRI \citep{u03,lebrun03}, covering the 20 -- 250 keV energy band, and from the two JEM-X monitors covering the 4 -- 22~keV energy range \citep{lund03}. The observation in revolution 1511 was performed in the hexagonal dithering mode, which permits to keep always the source within both the field-of-view of IBIS/ISGRI and JEM-X. For all other revolutions, we considered all the pointings for which the source was located at a maximum off-set angle with respect to the satellite aim point of $\lesssim12\fdg0$ for IBIS/ISGRI and $\lesssim 2\fdg5$ for JEM-X in order to minimize calibration uncertainties. The data reduction was performed using the standard {\sc offline science analysis (OSA)} version 10.2 distributed by the ISDC \citep{c03}. The algorithms used for the spatial and spectral analysis of ISGRI and JEM-X are described in \citet{gold03} and \citet{lund03}, respectively.  

\begin{figure}[h] 
\includegraphics[scale=0.44]{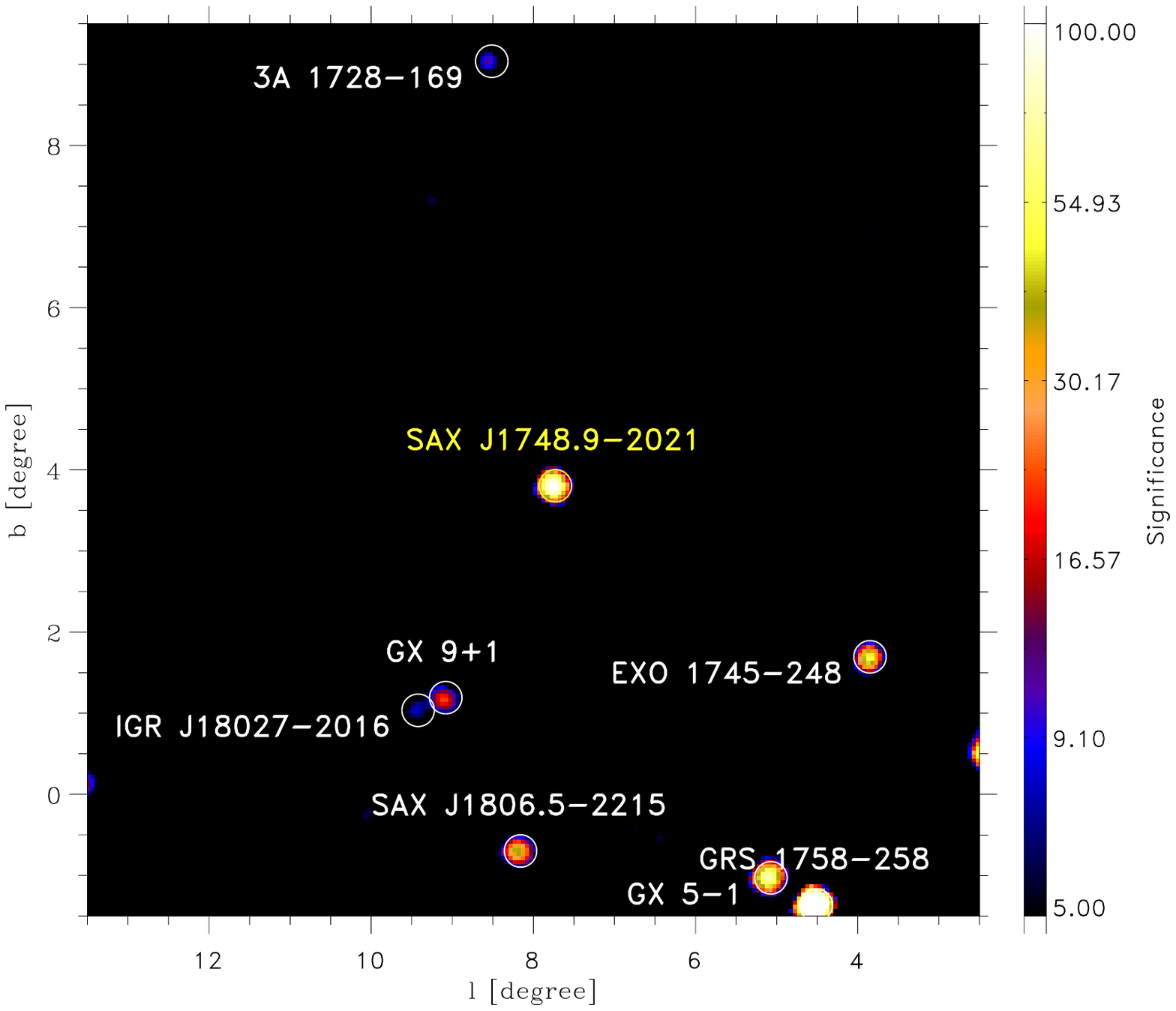}

%\centerline{\psfig{figure=MOSA.pdf, trim=4.5cm 6cm 3cm 8.8cm, scale=0.52}}

\caption{The IBIS/ISGRI sky image centered on \sax\ in the 20 -- 100 keV range of the  $\sim100$ ks ToO observation. The size of each pixel is 3$\arcmin$. }
\label{fig:mosa}
\end{figure}

In Fig. \ref{fig:mosa}, we show part of the ISGRI field of view (significance map) centered on the position of \sax. The source is clearly detected in the mosaic, and we estimated a detection significance of $\sim 126\ \sigma$ in the 20 -- 100~keV energy range. The best determined position is  $\alpha_{\rm J2000} = 17^{\rm h} 48^{\rm m} 52\fs20$ and $\delta_{\rm J2000} = -20{\degr}21\arcmin 32\farcs62$ with an associated uncertainty of $0\farcm4$ at 90\% confidence level  \citep{gros03}, consistent with the  best determined position \citep{pooley02}. 

Using the ISGRI data in the energy band 15 -- 300 keV, starting from revolution 1508, we performed a timing analysis to search for the pulsations. We applied the same timing techniques as used for the ISGRI data in the past \citep[see e.g.,][]{Kuiper03,mfb05,mfc07,falanga11,falanga12,defalco17a,defalco17b}, adopting the ephemeris reported in \citet{Sanna16}. No pulsed emission has been detected above 15 keV, which is in line with the very low pulsed fraction
of $\sim$2\% as detected by \xmm\ above 10 keV \citep{Sanna16}. %\red{JP: XMM does not have any response above 10 keV; how can one make any conclusions?}

We extracted the IBIS/ISGRI and JEM-X light curve of \sax\ at a time scale of typically $\sim 2-3$\,ks, i.e. the duration of a pointing  (science window).  To search for type-I X-ray bursts, we used the JEM-X data also at a time resolution of 2 seconds, as discussed in Sect.~\ref{sec:bursts}.  
The JEM-X and ISGRI spectra were extracted for the hard and soft state quasi-contemporaneous with \swift\ observations (see Sect.~\ref{sec:spe}). 

\begin{figure*}
\centering
%\centerline{\psfig{figure=Fig1, trim=1.5cm 6.5cm 1.3cm 6.2cm, scale=.43}}
\includegraphics[angle=0, scale=0.5]{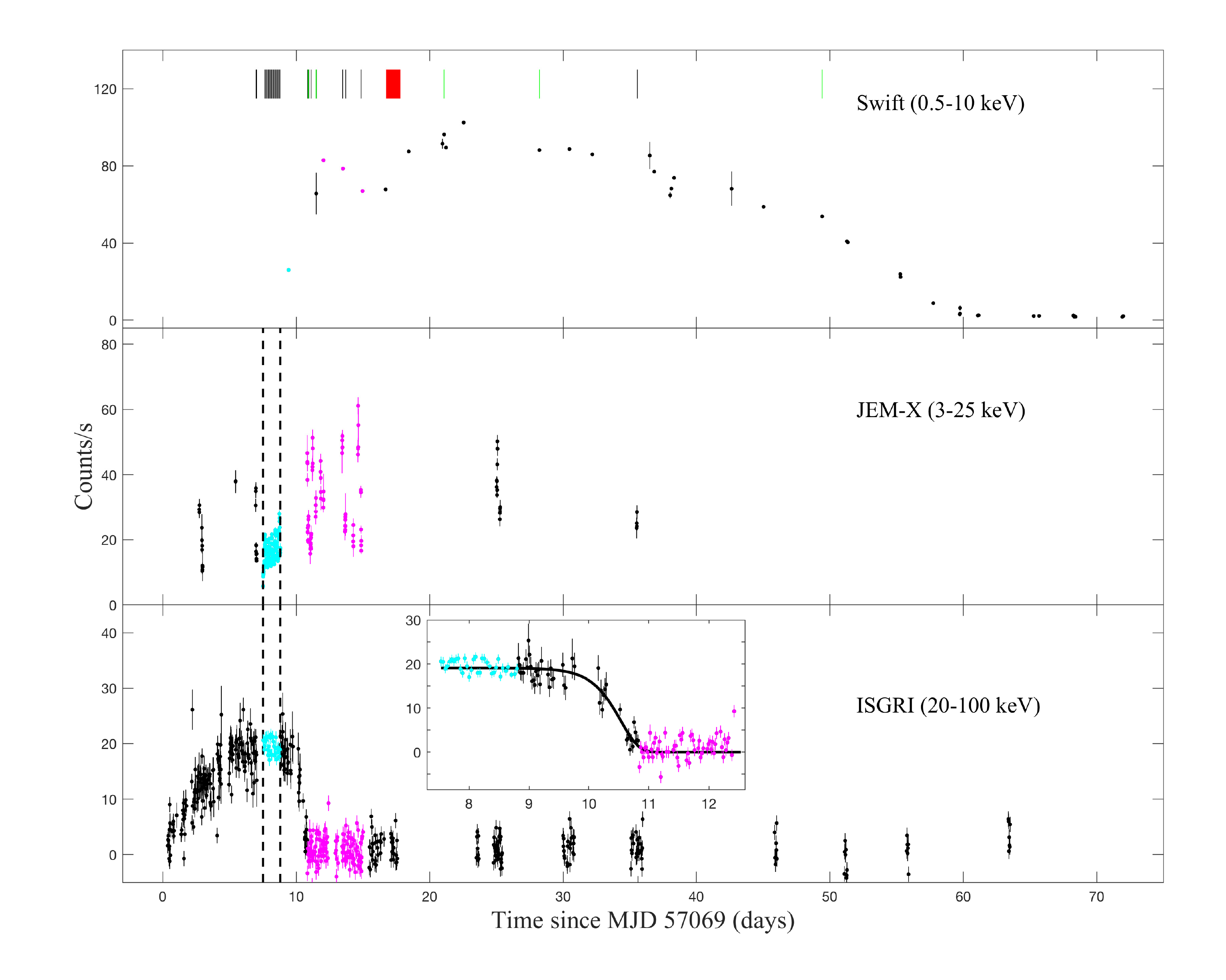}
\caption{\sax\ light curve during the 2015 outburst.  The vertical dashed lines,  indicate the $\sim100$ ks \Integ\ ToO time interval. The black, green, and red vertical lines indicate the times of the detected type-I X-ray bursts with \Integ, \swift\ and \xmm, respectively. In the bottom panel the inset shows  the hard ISGRI light curve zoomed in at the spectral change time interval. The count rate drops to zero within 0.5 day. The hard and soft state data points used in the spectral analysis are color coded with light blue and pink, respectively (see Sect. \ref{sec:spe}).}
\label{fig:lc}
\end{figure*} 
 
\subsection{Swift} 
\label{sec:swift}  

\swift/XRT \citep{burrows05} detected \sax\ in outburst after a 2 ks observation on 2015 February 18 \citep{Bozzo15}. The XRT monitoring campaign covered the 2015 outburst from February 25 to July 30 comprising a total of 40 pointings and the effective exposure time of $\sim43$~ks.

We processed the \swift/XRT data by using standard procedures \citep{burrows05} and the calibration files version 20160113. The \swift/XRT data were taken both in window-timing (WT) and photon-counting (PC) modes (processed with the {\sc xrtpipeline} v.0.13.2). Filtering and screening criteria were applied by using {\sc ftools} contained in the {\sc heasoft} software package (v6.19)\footnote{http://heasarc.gsfc.nasa.gov/docs/software.html.}. We extracted source and background light curves and spectra by selecting event grades of 0 -- 2 and 0 -- 12 for the WT and PC mode, respectively. We used the latest spectral redistribution matrices in the HEASARC calibration database. Ancillary response files, accounting for different extraction regions, vignetting and PSF corrections, were generated by using the {\sc xrtmarkf} task. When required, we corrected PC data for pile-up, and used the {\sc xrtlccorr} task to account for this correction in the background-subtracted light curve.

\subsection{XMM-Netwon}
\label{sec:xmm}

\sax\ was observed with \xmm\ \citep{jansen01} on March 4, 2015  (Obs.ID. 0748391301) continuously for about $\sim$100 ks with the first results reported in \citet{Pintore16}. We reanalysed the \xmm\ observation to obtain each burst flux and its fluence. Following the standard procedure, the TIMING data from European Photon Imaging Camera (EPIC)-pn \citep{Struder01} were reduced and the last 5 ks of the data were ignored because of flaring background. The data were filtered with FLAG=0 and PATTERN$\leq$4. After the extraction of the lightcurve, for each burst, we located its start and end time at which the \textbf{count rate} increased and decreased to 10\% of its peak value above the persistent intensity level, respectively. The spectrum of each burst was obtained in the range RAWX=[31:43] without the column RAWX=37 to reduce the pile-up effect. The persistent spectrum prior to a burst was regarded as its background. Each burst spectrum was fitted by the model {\sc tbabs*bbodyrad}, and the burst bolometric flux was calculated using the relation $F=1.076\times10^{-11}(kT_{\rm bb}/1~{\rm keV})^4K_{\rm bb}~{\rm ergs~cm^{-2}~s^{-1}}$\citep{Galloway08}, where $kT_{\rm bb}$ and $K_{\rm bb}$ are the blackbody temperature and normalization, respectively. The {\em XMM-Newton} data reduction and analysis was similar to the procedure described in \citet{Pintore16}. All reported fluxes are unabsorbed.

%We reanalysed the \xmm ~observation reported by \citet{Pintore16}. Following the standard procedure,  the TIMING data from European Photon Imaging Camera (EPIC)-pn were reduced and the last 5 ks of the data were ignored because of flaring background. The data were filtered with FLAG=0 and PATTERN$\leq$4. After extracted the light curve, we located the start and stop time of each burst. The spectrum of each burst was obtained in the range RAWX=[31:43] without the column RAWX=37 to reduce the pile-up effect. The persistent spectrum prior to a burst was regarded as its background. Each burst spectrum was fitted by the model {\sc tbabs*bbodyrad}, and the burst flux in the range 0.1-50 keV were calculated by the command {\sc cflux}. %We also extracted the sky and detector background from the column RAWX=[3:5]  to model the persistent emission.  , and the persistent spectrum was adopted the model from  \citet{Pintore16}. The burst and persistent flux in the range 0.1-50 keV were calculated by the command cflux. 

\section{Outburst light curve} 
\label{sec:lc} 

We report in Fig.~\ref{fig:lc} the outburst profile of \sax\ from February 16 to May 2, 2015. \sax\ exhibited significant spectral variation during the outburst, passing from  hard to soft state within $\approx 0.5$ d around MJD 57079.5. The hard state lasted for the first $\approx10$ d, while the soft state lasted about 50~d, until the source returned to quiescence. Most AMXPs that underwent an outburst for a few weeks to months showed a common outburst profile, i.e., the light curve decays exponentially until it reaches a break, after which the flux drops linearly to the quiescence level without exhibiting strong spectral variation \citep[e.g.,][]{gdb02,gp05,mfa05,mfb05,mfc07,ip09,falanga11,falanga12,defalco17a,defalco17b}. However, this source is known to be an atoll source with significant flux variations between hard and soft state and showing intermittent pulsations whose origin is unclear \citep[see e.g.,][]{Patruno09}.

\section{Spectral analysis} 
\label{sec:spe}

The spectral analysis was carried out using {\sc xspec} version 12.6 \citep{arnaud96}. We studied the broad-band X-ray spectrum divided in the hard and soft state, see Fig. \ref{fig:lc}. For the hard state we consider the \Integ/JEM-X (3.5 -- 22) keV and ISGRI (20 -- 250) keV ToO data between MJD 57076.52 -- 57077.82, and the quasi-simultaneous \swift/XRT/WT (1 -- 10) keV data  starting on MJD 57078.43  for 1~ks (obs. ID 00033646003). For the soft state we consider the \Integ/JEM-X (4 -- 22) keV and ISGRI (20 -- 50) keV public data between MJD 57079.83 -- 57084.03 and one simultaneous \swift/XRT/WT  (1 -- 10) keV data for a total of 2 ks (obs ID. 00033646004). We did not merge together the additional two close-by \swift\ pointings ID.~00033646006 and ID.~00033646005, as the roll angle between the three pointings was different and the statistical quality of the first pointing we used was sufficient to perform an accurate broad-band fit. We verified a posteriori that performing the same broad-band fit with the other two \swift\ pointings would have resulted in similar values of the model parameters (to within the uncertainties). 
For all spectra the bursts intervals have been removed from the data set. For each instrument, a multiplication factor was included in the fit to take into account the uncertainty in the cross-calibration of the instruments. For all fits the factor was fixed at 1 for the \Integ/ISGRI data. All uncertainties in the spectral parameters are given at a $1\sigma$ confidence level for a single parameter.

\begin{figure}[htb] 
\centering 
\includegraphics[scale=0.33,angle=0 ]{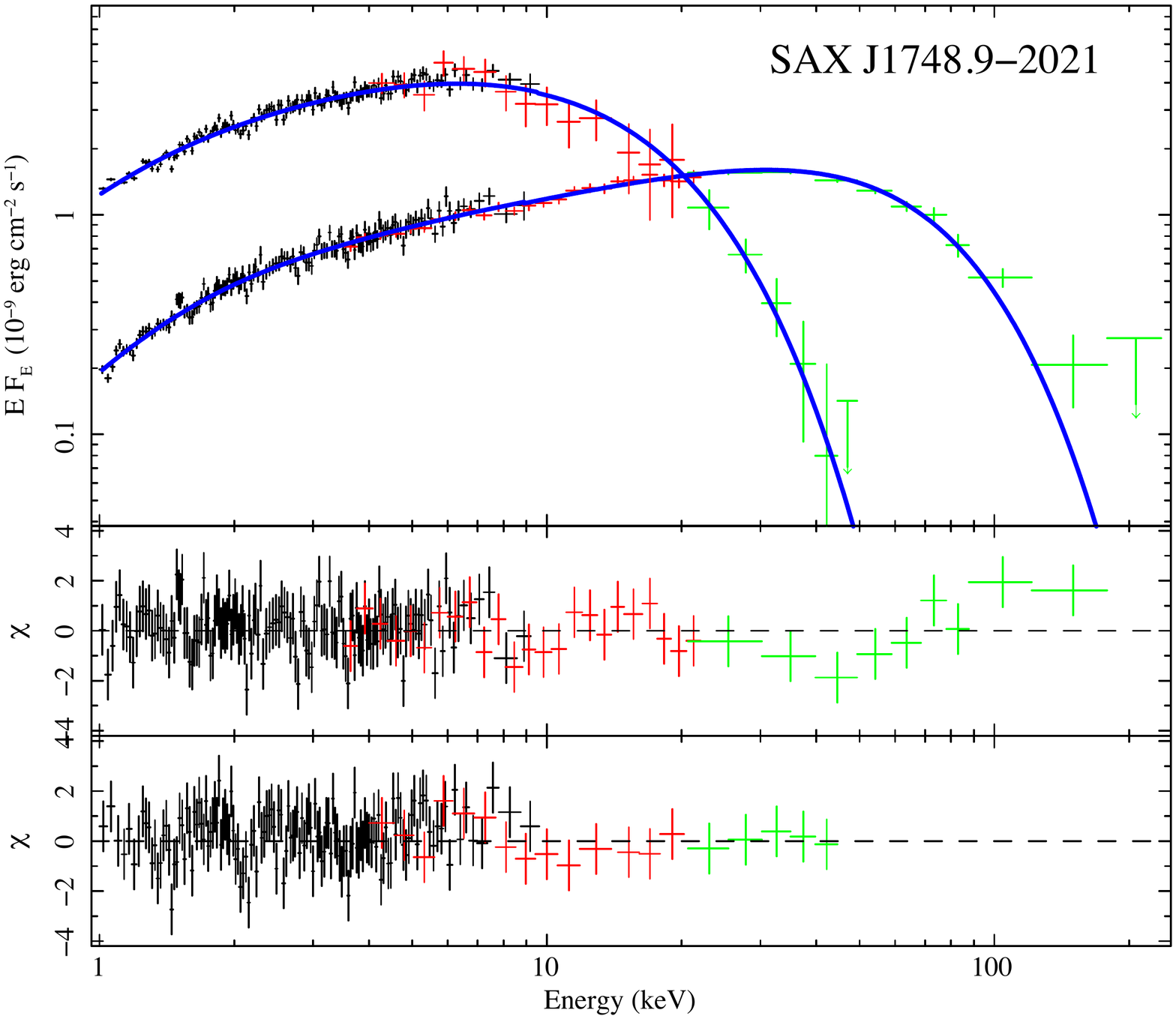}
\caption{The unfolded broad-band spectrum of \sax\ showing the hard and soft spectral states. The data points are from the \swift/XRT (black crosses), \Integ/JEM-X (red crosses), and \Integ/ISGRI (green crosses). The best fit is obtained with the \compps\ model, represented by a solid blue line. The lower panels show the residuals between the data and the model. The \swift/XRT spectra have been rebinned for a better visualization. }%\red{JP: make blue line thicker}
\label{fig:spe} 
\end{figure} 

We fit all the combined XRT/JEM-X/ISGRI average spectra separately for the  hard and soft spectral state, respectively, using first a simple and phenomenological  cutoff power-law model and afterwards a physical motivated thermal Componization model, {\sc compps}, in the slab geometry \citep{ps96}. The  later model has been used previously to fit AMXPs broad-band spectra  \citep[e.g.,][]{gdb02,gp05,mfa05,mfb05,mfc07,ip09,falanga11,falanga12,defalco17a,defalco17b}. The main parameters are the absorption column density,   $N_{\rm H}$, the Thomson optical depth, $\tau_{\rm T}$, across the slab, the electron temperature, $kT_{\rm e}$, the temperature, $kT_{\rm seed}$, of the soft-seed thermal photons (assumed to be injected from the bottom of the slab), the apparent area of the seed photons, $A_{\rm seed}$, and $\cos \theta $ that we fixed at 0.5, where $\theta$ is the inclination angle between the slab normal and the observer line of sight. The best-fit parameters of all models are reported in Table \ref{tab:spec}.  At variance with \citet{Pintore16} we could not find any evidence of the {\sc diskbb} component from the residuals of our best fit comptonization models (see also Fig.~\ref{fig:spe}).  

%We noticed, that the values of the normalization constants were all in the range $1.0\pm0.3$, as expected for quasi-simultaneously average spectral fits. 
 
We note that the column density between the two states changes from $N_{\rm H}\sim0.5 \times 10^{22}~ {\rm cm}^{-2}$ to $N_{\rm H}\sim0.8 \times 10^{22} ~{\rm cm}^{-2}$, as well as the plasma temperature, $kT_{\rm e}$, and the optical depth, $\tau_T$ (see Table \ref{tab:spec}). It is remarkable how this spectral change from a hard to a soft state occurred within a half day (see Fig. \ref{fig:lc}). 

\begin{table}[htb] 
\caption{Best fit parameters determined for the hard and soft spectral state of \sax. }
\centering
\scalebox{0.755}{
\begin{tabular}{lllll} 
\hline
\hline 
& \multicolumn{2}{c}{Hard state}  & \multicolumn{2}{c}{Soft state} \\
& {\sc cutoffpl}&  \compps & {\sc cutoffpl}&  \compps\\
\hline 
\noalign{\smallskip}  
$N_{\rm H}\ (10^{22} {\rm cm}^{-2})$ & $0.57\pm0.03$ & $0.52\pm0.08$ &$0.90\pm0.06$ & $0.8\pm0.1$\\ 
$\Gamma$ & $1.37\pm0.03$&  -- &$1.1\pm0.1$ & -- \\
$E_{\rm cut}$ & $40\pm2$ & -- & $6.7\pm1.4$ & --\\
$kT_{\rm e}$ (keV)& -- & $18.1\pm0.8$ & -- & $5.0\pm0.5$\\ 
$kT_{\rm seed}$ (keV)& -- & $0.39\pm0.06$& --  & $0.28\pm0.15$ \\
$\tau_{\rm T}$ & -- &  $2.9\pm0.1$ & --  & $3.8\pm0.5$ \\ 
%$\cos \theta $ & -- &  0.6 (fix) & -- &  0.6 (fix)\\
$A_{\rm seed}$  & -- &  $5000^{+1450}_{-1300}$ &--& $10300^{+6000}_{-6000}$\\ 
$F_{\rm bol}$ (10$^{-9}$ cm$^{-2}$ s$^{-1}$)$^{a}$ & $3.1\pm0.2$& $3.1\pm0.2$ & $6.1\pm0.2$ & $6.1\pm0.2$\\
$\chi^{2}_{\rm red}/{\rm dof}$ & 1.05/516 & 1.06/515 & 0.98/579  & 0.98/578 \\
\noalign{\smallskip}  
\hline  
\end{tabular}} 
\tablefoot{ \tablefoottext{a}{Unabsorbed flux in the 0.1 -- 250 keV and 0.1 -- 50 keV energy range for the hard and soft states, respectively.
For the  {\sc cutoffpl} model, the parameters are the photon index, $\Gamma$, the cutoff energy, $E_{\rm cut}$. For the \compps\ model, the main parameters are the Thomson optical depth, $\tau_{\rm T}$, across the slab, the electron temperature, $kT_{\rm e}$, the temperature, $kT_{\rm seed}$, of the soft-seed thermal photons and the apparent area of the seed photons, $A_{\rm seed}$. 
 }}
\label{tab:spec} 
\end{table} 

%\begin{table}[h] 
%\caption{Best fit parameters determined for the soft spectral state of \sax.}
%\centering
%\scalebox{0.8}{
%\begin{tabular}{llll} 
%\hline 
%& {\sc cutoffpl} & {\sc comptt} & \compps \\
%\hline 
%\noalign{\smallskip}  
%$N_{\rm H}\ (10^{22} {\rm cm}^{-2})$ & $0.90\pm0.06$ & $0.60\pm0.1$ & $0.8\pm0.1$\\ 
%$\Gamma$ & $1.1\pm0.1$ & -- & --\\
%$E_{\rm cut}$ & $6.7\pm1.4$ & -- & --\\
%$kT_{\rm e}$ (keV)& -- & $2.5\pm0.2$ & $5.0\pm0.5$ \\ 
%$kT_{\rm seed}$ (keV)& --  & $0.44\pm0.05$ & $0.28\pm0.15$\\ 
%$\tau_{\rm T}$ & -- &  $7.8\pm0.7$ & $3.8\pm0.5$ \\
%$A_{\rm seed}$ & -- & -- & $10314\pm6000$ \\
%$\cos \theta $ & -- & -- &  0.60 (fix) \\
%$\chi^{2}_{\rm red}/{\rm dof}$ & 0.98/579  & 0.98/578 & 0.98/578 \\
%\noalign{\smallskip}  
%\hline  
%\end{tabular}}  
%\label{tab:spec2} 
%\end{table} 

\section{The type-I X-ray bursts} 
\label{sec:bursts} 

\sax\ is also known to be a burster source, as during its 2015 outburst 56 type-I X-ray bursts have been detected (see Fig.~\ref{fig:lc}), of which 26 found by \Integ/JEM-X, 5 by \swift, and for completeness we added also 25 bursts observed by \xmm\ and reported by \citet[][]{Pintore16}. 
These bursts occurred during both the hard and soft states. However, we  analyzed in more detail the bursts detected during the continuous \Integ\ ToO and \xmm\ observations, where the bursts were separated by two distinct time intervals of around $\Delta t_{\rm rec, hard}\sim2$ hr (hard state, \Integ\ data)  and $\Delta t_{\rm rec, soft}\sim1$ hr (soft state, \xmm\ data). All bursts lasted on average $\approx100$ ~s and showed a rise time of $\sim2-3$~s. The burst decay profile could be well fitted with an exponential function and the correspondingly derived e-folding time is $\tau_{\rm fit}=23\pm3$ s, that is in agreement with the value reported by \citet[][]{Pintore16}, estimated to be $25\pm 1$ s (typo in their article, private communication). We note, that all the bursts outside these two observations also showed comparable $\tau_{\rm fit}$ values and similar burst profiles.

\begin{figure}[htb] 
\centering 
 \includegraphics[scale=0.45]{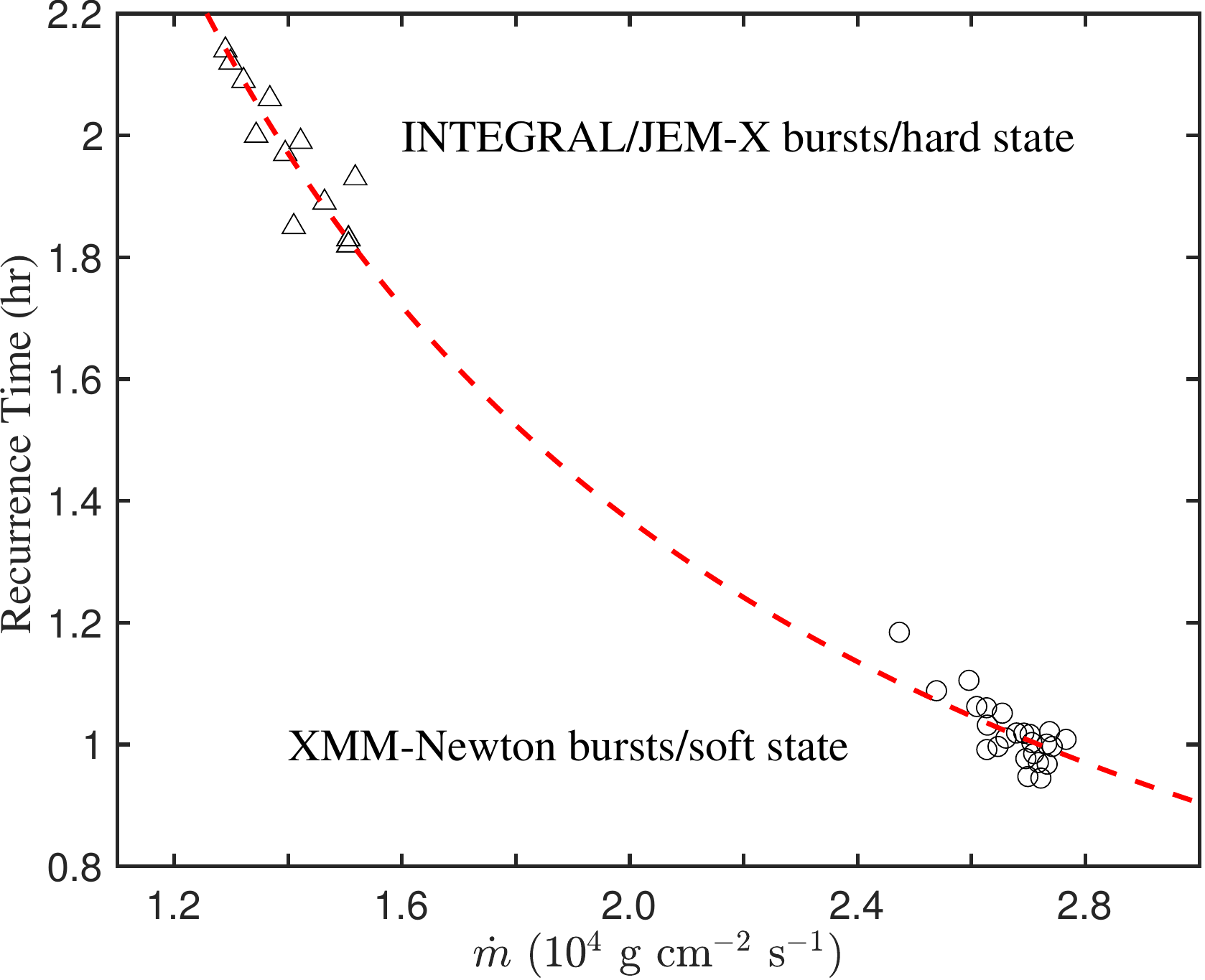}

\caption{Recurrence time versus local mass accretion rate per unit area onto the compact object, where the circles and triangles represent the type-I X-ray bursts detected by \xmm\ and \Integ/JEM-X, respectively. 
 The dashed red lines represent the best-fit power-law model for all bursts $\Delta t_{\rm rec} \sim \langle  \dot{m} \rangle^{-1.02\pm0.03}$. The recurrence time decreased from $\sim$ 2 hr to $\sim$ 1 hr as the persistent spectrum showed the transition from the hard to the soft state.%The fuel burned by the bursts marked by triangles and circles  is  mixed hydrogen/helium.
}
\label{fig:Fig4} 
\end{figure} 
 
The averaged flux in the 0.1 -- 50 keV and 0.1 -- 250 keV energy range of the persistent emission during the soft and hard states, respectively,  was $F_{\rm pers, soft}\approx(5.48-6.13)\times10^{-9}$  erg cm$^{-2}$ s$^{-1}$ for \xmm\ bursts and $F_{\rm pers, hard}\approx(2.88-3.35)\times10^{-9}$  erg cm$^{-2}$ s$^{-1}$ when \Integ/JEM-X detected the bursts. The corresponding luminosities are $L_{\rm pers, soft}\approx(4.4-4.9)\times 10^{37}$ erg s${}^{-1}$, i.e. $(12-13)\%\ L_{\rm Edd}$, and $L_{\rm pers, hard}\approx(2.32-2.70)\times 10^{37}$ erg s${}^{-1}$, i.e. $(6-7)\%\ L_{\rm Edd}$, where $L_{\rm Edd} = 3.8\times10^{38}$ erg cm$^{-2}$ s$^{-1}$ is the Eddington luminosity adopted by \citet{Kuulkers03}. We have used as distance to the source $d=8.2$ kpc. 

The local accretion rate per unit area onto the compact object is $\dot{m} = L_{\rm pers} (1+z) (4\pi R^2(GM/R))^{-1}$, i.e., $\dot{m}_{\rm soft} \approx 2.67\times10^4$ g cm$^{-2}$ s$^{-1}$ and $\dot{m}_{\rm hard} \approx (1.30-1.51)\times10^4$ g cm$^{-2}$ s$^{-1}$, where the gravitational redshift is $1+ z = 1.31$ for a canonical NS with a mass $M=1.4M_\odot$ and a radius of $R=10$ km. The observed recurrence time increases with $\dot{m}$ as $\Delta t_{\rm rec} \sim \langle  \dot{m} \rangle^{-1.02\pm0.03}$ (see Fig.~{\ref {fig:Fig4}}).

For the bursts in the soft state, the total burst fluence is $f_{\rm b, soft}\approx0.35\times10^{-6}$ erg cm$^{-2}$, obtained from the burst flux multiplied by the corresponding burst duration. For the bursts in the hard state, the peak flux is $F_{\rm peak, hard}\approx1.6\times10^{-8}$  erg cm$^{-2}$ s$^{-1}$, that multiplied by $\tau_{\rm fit}\approx23$ s give an estimation of the total burst fluence of $f_{\rm b, hard}\approx0.37\times10^{-6}$ erg cm$^{-2}$. We note, that these values are close to the directly measured total fluence per burst. 

%The mean peak fluxes are $F_{\rm peak, soft}\approx2.3\times10^{-8}$  erg cm$^{-2}$ s$^{-1}$ and

We can then estimate the ignition depth at the onset of the burst with the equation $y_{\rm ign} =4\pi f_{\rm b}\, d^2\, (1+z)\,(4\pi R^2Q_{\rm nuc})^{-1}$, where the nuclear energy generated for solar composition (assuming $\bar X=0.7$) is $Q_{\rm nuc}\approx 1.31+6.95\bar X-1.92\bar X^2~ {\rm MeV/nucleon}\approx4.98\ {\rm MeV/nucleon}$ (Goodwin et al. submitted). Note that the considered values $Q_{\rm nuc}$ include losses owing to neutrino emission as determined by using KEPLER. Therefore we obtained $y_{\rm ign, soft}\approx0.59\times10^{8}$ g cm$^{-2}$  and $y_{\rm ign, hard}\approx0.63\times10^{8}$ g cm$^{-2}$ for the soft and hard state, respectively.

%pure helium (assuming a mean hydrogen mass fraction at ignition $\bar X=0$) is $Q_{\rm nuc}\approx 1.31+6.95\bar X-1.92\bar X^2~ {\rm MeV/nucleon}\approx1.31\ {\rm MeV/nucleon}$  and for 

%$y_{\rm ign, soft}\approx 2.3\times10^{8}$ g cm$^{-2}$  for helium and 
 %$y_{\rm ign, hard}\approx2.5\times10^{8}$ g cm$^{-2}$ for helium and
%$\Delta t_{\rm rec, soft}\sim 3.1$~hr in case of helium burning, while
%$\Delta t_{\rm rec, hard}\sim 6.5$~hr in case of helium burning, while

Once the ignition depth is known, the recurrence time between the bursts can be calculated by using the equation $\Delta t_{\rm rec} =(y_{\rm ign} /\dot{m})(1 + z)$. For the soft state, we obtain 
 $\Delta t_{\rm rec, soft}\sim 0.81$~hr, and for the hard state we obtain  $\Delta t_{\rm rec, hard}\sim 1.6$~hr in the mixed hydrogen/helium burning case. For the pure helium burning, the predicted recurrence time is much longer.  Compared with the observed recurrence time and $\dot{m}$, these results confirm that the burst type is mixed hydrogen/helium with solar abundances for both the hard and soft state (see Fig. \ref{fig:Fig4}). In addition this can be additionally validated by the $\alpha$ value \citep[see e.g.,][]{strohmayer06}. We computed the ratio of the integrated persistent flux to the burst fluence, $\alpha=\tau_{\rm rec}F_{\rm per}/f_{\rm b}$, which are  \textbf{$64\pm10$} for the soft bursts and $60\pm6$ for the hard bursts. The $\alpha$-values are consistent with the mixed hydrogen/helium bursts from \sax ~reported by \citet{Galloway08}.

If the anisotropy of the persistent and burst emission, i.e., with a factor of $\xi_{\rm p}$ and $\xi_{\rm b}$ respectively, are considered, it will introduce several consequences \citep[see][and references therein]{He16}. The local accretion rate ($\dot{m}$) should be multiplied by the factor $\xi_{\rm p}$, however, it will not effect the obtained power-law index of the $\Delta t_{\rm rec}-\langle \dot{m}\rangle $ relation if the accretion disk shape did not change dramatically from the soft to hard transition (in both cases, $\dot{m}$ is multiplied by the same factor). The predicted recurrence time is proportional to the observed value, $f_{\rm b}/F_{\rm pers}$, so it should be changed with a factor of $\xi_{\rm b}/\xi_{\rm p}$. The intrinsic $\alpha$ parameter  can be obtained from the observed value divided by a factor of $\xi_{\rm b}/\xi_{\rm p}$ \citep{Fujimoto88}. The $\xi_{\rm b}/\xi_{\rm p}$ factor is estimated in the range 1 -- 1.5 for the low inclination system of \sax\ \citep{He16}. These modifications still support the mixed hydrogen/helium bursts occurred in \sax.

\section{Discussions and Conclusions} 

\sax\ has shown 17 bursts during the 1998 outburst detected by \BS, 16 bursts during the 2001 outburst, 2 bursts during the 2005 outburst, and 13 bursts during the 2009-2010 outburst, all detected by \rxte\ \citep{Galloway08}. 
The total number of X-ray bursts was enlarged by the 56 bursts detected during the 2015 outburst. The bursts in soft and hard states are similar in terms of their e-folding time, the peak flux, and the total fluence. Compared with the calculated recurrence time, we conclude that all bursts are generated from  mixed  hydrogen/helium burning on the NS surface. However, this source appeared to alternate between helium or mixed  hydrogen/helium bursts independently of the persistent flux \citep{Galloway08}, while during the 2015 outburst, studied in detail in this work, the mixed hydrogen/helium bursts follow the theoretical predicted recurrence time as function of the mass accretion rate $\Delta t_{\rm rec} \sim \langle  \dot{m} \rangle^{-1.02\pm0.03}$ (see Fig.~{\ref {fig:Fig4}}). 

We have studied the spectral behaviour of \sax\ during its 2015 outburst by using the available \Integ, \swift, and \xmm\ data. The source shows a spectral change from hard to soft state within $\sim0.5$ day around MJD 57079.5  (see Fig.~\ref{fig:lc}). 
%The energy spectra of both states are well fitted by a thermal Comptonization model in slab geometry (see Fig.~\ref{fig:spe}). The parameters of this model confirms adequately this spectral change, since we have that in the hard state the best-fit parameters are $N_{\rm H}\sim0.5\times10^{22}~{\rm cm^{-2}}$, $kT_{\rm e}\sim18.1$ keV, and $\tau\sim 2.9$; while in the soft state $N_{\rm H}\sim0.8\times10^{22}~{\rm cm^{-2}}$, $kT_{\rm e}\sim5.0$ keV, and $\tau\sim3.8$ (see Table~\ref{tab:spec}).
Usually, the hard to soft transitions occurred in the rise phase of the outbursts in LMXBs \citep[see][for reviews]{Remillard06}. The hard to soft transitions in LMXBs were explained in different ways, i.e., the increasing of mass accretion rate \citep{Esin97}, the shrinking of the corona size \citep{Homan01}, the competition between inner halo and outer standard thin accretion disk \citep{Smith02}, and non-stationary accretion \citep{Yu09}.  The duration of the outbursts' rise phase in LMXBs can be regarded as the timescale of the hard to soft transition, which has a mean value of 5 days and a minimum value of $\sim$ 1 day \citep{Yan15}.
%Over the past few decades, another four outbursts have been observed in \sax, two of them (the outbursts in 2005 and 2009$-$2010) showed soft to hard spectral transition, the outburst in 2001 exhibited hard to soft spectral transition, while the outburst in 1998 only performed as hard state. The spectral changes (soft to hard or hard to soft) in LMXBs are probably caused by the accretion disk instabilities. 
%\citet{Yan15} measured the e-folding rise timescale of outbursts in LMXBs and found that the 
Compared with the large sample of outbursts in 36 LMXBs \citep{Yan15}, the rapid spectral transition in \sax\ during the 2015 outburst had the shortest duration.  We suggest that the production of soft X-ray photons from type-I X-ray bursts may have accelerated the hard-to-soft spectral transition. 

Before the transition, the persistent emission in the 20 -- 100 keV range maintained at a high level in $\sim$ 5 days (see the bottom panel in Fig.~\ref{fig:lc}). Considering the recurrence time of $\sim$2 hrs in the hard state, we expect that a total number of type-I bursts of $\sim$ 60 should have occured in SAX J1748.9-2021, and so $\sim$ 35 of these are probably missed because the source was outside the field of view of JEM-X. With the averaged energy release, $3\times10^{39}~{\rm ergs}$, of type-I X-ray bursts, the total energy of $1.8\times10^{41}~{\rm ergs}$ was emitted  in the hard state.  The X-ray bursts in \sax\ may play two roles in the hard-to-soft transition by ejecting a bunch of soft X-ray photons and dragging the accretion flow inward to enhance the persistent X-ray emission \citep{walker92}.  In both cases, the produced soft photons pass through the corona and cool down the electrons there, which can reduce the corona size and shorten the duration of the hard-to-soft transition. Moreover, we found  the source EXO 1745-248 showing a similar behavior, especially considering the two outbursts at MJD 51776 and MJD 52459, respectively. The first outburst was more energetic than the second one. However, the hard to soft transition duration in the first outburst, i.e., $\sim$ 4 days, was smaller than the second one, i.e., $\sim$ 5 days \citep{Yan15}. Meanwhile, dozens of X-ray bursts happened before the hard to state transition in the first outburst \citep{Galloway08}. Considering that more total energy released in the first outburst, we suggested that X-ray bursts may also accelerate the hard to soft spectral transition during the first outburst in EXO 1748-248.

%This finding two phase type-I X-ray bursts have been already noticed by \citet{Galloway08}, by comparing different type-I X-ray bursts during different outbursts. %However, here we observed in one outburst the change of burst type.  

\label{sec:discussion} 

\begin{acknowledgements} 
We appreciate the suggestions from the referee, which improve the paper. Z.L. thank S.S. Weng for helpful discussions. Z.L. was supported by the Swiss Government Excellence Scholarships for Foreign Scholars. Z.L. thanks the International Space Science Institute and University of Bern for the hospitality. This work was supported by National Natural Science Foundation of China (Grant No. 11703021) and Hunan Provincial Natural Science Foundation of China (Grant Nos. 2017JJ3310 and 2018JJ3495). V.D.F. and M.F. acknowledge the Swiss National Science Foundation project 200021\_149865, who financed this research project. V.D.F. acknowledges the International Space Science Institute in Bern for their support.
J.P. was supported by the grant 14.W03.31.0021 of the  Ministry of Education and Science of the Russian Federation.

\end{acknowledgements} 

%\bibliographystyle{aa}
%\bibliography{references}

\end{document}